\documentclass[aps,prper,longbibliography,reprint,graphicx,floatfix]{revtex4-1}

\usepackage{graphicx}
\usepackage{placeins}

\draft 

\newcommand{\etal}{et~al.}

\begin{document}

\title{Major Outcomes of an Authentic Astronomy Research Experience Professional Development Program: An Analysis of 8 Years of Data from NITARP}

\author{L. M. Rebull}
\email{rebull@ipac.caltech.edu}
\affiliation{Caltech-IPAC/IRSA, SSC, and NITARP, 1200 E. California Blvd., MS 100-22, Pasadena, CA 91125 USA}

\author{D. A. French}
\affiliation{Wilkes University, 84 West South St., Wilkes-Barre, PA 18766 USA}

\author{W. Laurence}
\affiliation{Create-osity, 3187 Julies Dr., Park City, UT 84098 USA}

\author{T. Roberts}
\affiliation{Caltech-IPAC/ICE, 1200 E. California
Blvd., MS 314-6, Pasadena, CA 91125 USA}

\author{M. T. Fitzgerald}
\affiliation{Edith Cowan Institute for Education Research, 270 Joondalup Dve, Joondalup WA 6027, Australia}

\author{V. Gorjian}
\affiliation{NASA/JPL, 4800 Oak Grove Dr., Pasadena, CA 91101 USA}

\author{G. K. Squires}
\affiliation{Caltech-IPAC/ICE, 1200 E. California Blvd., MS 314-6, Pasadena, CA 91125 USA}

\date{\today}

\begin{abstract}
The NASA/IPAC Teacher Archive Research Program (NITARP) provides a
year-long authentic astronomy research project by partnering a
research astronomer with small groups of educators. NITARP has worked
with a total of 103 educators since 2005. In this paper, surveys are
explored that were obtained from 74 different educators, at up to four
waypoints during the course of 13 months, from the class of 2010
through the class of 2017; those surveys reveal how educator
participants describe the major changes and outcomes in themselves
fostered by NITARP. Three-quarters of the educators self-report some
or major changes in their understanding of the nature of science.  The
program provides educators with experience collaborating with
astronomers and other educators, and forges a strong link to the
astronomical research community; the NITARP community of practice
encourages and reinforces these linkages. During the experience,
educators get comfortable with learning complex new concepts, with
$\sim$40\% noting in their surveys that their approach to learning has
changed. Educators are provided opportunities for professional growth;
at least 12\% have changed career paths substantially in part due to
the program, and 14\% report that the experience was ``life
changing.''  At least 60\% express a desire to include richer, more
authentic science activities in their classrooms.  This work
illuminates what benefits the program brings to its participants; the
NITARP approach could be mirrored in similar professional development
(PD) programs in other STEM subjects. 
\end{abstract}


\maketitle 

\section{Introduction}
\label{sec:intro}

Science education has been in an on-going state of reform
\cite{deboer2014history}.  Previously, the emphasis was on
incorporating inquiry-based pedagogies in the classroom
\cite{deboer2014history}.  With the adoption of the Next Generation
Science Standards (NGSS; \cite{standards2013next}) by many states,
current reform efforts call for K12 science education to incorporate
authentic scientific inquiry, which models the behaviors of practicing
scientists \cite{crawford2014,naframework}.  Incorporating authentic
scientific inquiry may be a daunting task, as educators may not have
had that experience themselves. Approximately two thirds of physics,
physical science, and earth science teachers do not have majors,
minors, or certifications in these areas, but are teaching this
content (Schools and Staffing Survey 2012 as cited in Marder
\cite{marder2017recruiting}); Ref.~\cite{nsfreferee} reports that
80-90\% of educators hold certifications in their subject, but that is
not the same as having an undergraduate major or minor in their field,
or for that matter science research experience. An informal survey of
a very small sample of 61 educators \cite{stalnaker2018} suggests that
about 40\% of educators did not have undergraduate research
experiences at all.  Professional development (PD) opportunities for
science teachers have been changing to meet the shifting demands in
the classroom.  Science PD programs have been moving away from the
one-day, ``sit and get" workshop formats to longer PD formats that
take place over several months to one year \cite{wei2009national}. 
Such formats have been shown to be more effective at enacting teacher
change \cite{yoon2007reviewing}. An example of the dozen or so such
programs available for educators in any physical or biological science
include the Research Experience for Teachers (RET) and Math and
Science Partnership (MSP) programs supported by the National Science
Foundation (NSF); Fitzgerald \etal~\cite{fitzgerald2014review}
includes several astronomy programs involving high school educators. 

Though the NGSS calls for teachers to incorporate more authentic
scientific inquiry in the classroom, the research shows teachers may
need additional support developing good, scientific research questions
and projects.  While teachers who have not conducted scientific
research may be able to identify flaws in others' experimental
designs, teachers may have difficulties developing their own
scientific research project \cite{taylor2003}.  Professional
development providing teachers with opportunities to do authentic
scientific research by closely collaborating with a research scientist
may provide the additional support teachers need in the NGSS era.  

A review of the literature yields promising results for PD
opportunities where teachers conduct research with scientists. Such
programs have been shown to increase teachers' content knowledge
\cite{dresner2006teacher,houseal2014impact,laursen2007good,raphael1999research}.
They have also been shown to increase teachers' knowledge and
confidence of using scientific and laboratory instruments and
techniques \cite{dresner2006teacher,westerlund2002}. 

Participating in a scientific research experience has led to changes
in the classroom.  After participating in scientific research,
teachers increased lab experiences for their students
\cite{westerlund2002}.  Teachers also incorporated more inquiry-based
pedagogies in their classroom
\cite{blanchard2009,houseal2014impact,laursen2007good,westerlund2002}.

Collaborating with a scientist was reported to positively change
teachers' views of science and scientists  \cite{houseal2014impact}.
Teachers reported being more enthusiastic about science and science
research \cite{westerlund2002}.  Teachers reported continuing the
collaboration with their mentor scientist
\cite{raphael1999research,westerlund2002}. Finally, teachers noted
their appreciation for opportunities for collaboration and support
from scientists \cite{dresner2006teacher,laursen2007good}. This
suggests that teachers are developing and sustaining a community from
participating in research experiences. 

The most substantial previous work on teacher research experiences
specifically in astronomy is Buxner's qualitative study
\cite{buxner2014exploring} of teachers' understanding about scientific
inquiry and the nature of science within three different summer
teacher research experience programs. Project durations in this study
ranged from one week to two months. Buxner found that the outcomes
intended by project directors, such as strong changes in teachers'
understanding of science inquiry and how to implement research with
their students, did not commonly occur, although smaller magnitude
improvements were noted. However, teachers self-reported significant
changes in their science pedagogy, personal and professional growth,
confidence and effective classroom activity. However, Buxner goes on
to note that ``systematic research across multiple programs using
similar protocols is highly limited.'' Additionally, when researching
the Students, Teachers, and Rangers, and Research Scientists (STaRRS)
Student-Teacher-Scientist Partnership, Houseal
\etal~\cite{houseal2014impact} found ``shifts in teachers' attitudes
regarding science and scientists, and shifts in their pedagogical
choices" (p. 84).  Research into teacher research experiences is also
lacking outside of astronomy. Sadler \etal~\cite{sadler2010} explore
53 studies of science research apprenticeship experiences, of which 11
were focused on teachers. They call for a greater methodological
diversity, to explore more ways of direct and valid measures of
outcome variables and also, most relevant to this particular study, is
to provide fine grained analyses of programmatic features to yield
additional insights.

NITARP, the NASA/IPAC Teacher Archive Research Program
(http://nitarp.ipac.caltech.edu), has worked with more than 100
teachers  over the last 10 years. Small groups of participant
educators are paired with a mentor astronomer and involved in a
year-long research project using professional astronomy tools and
archival data. The teams present their  results at the American
Astronomical Society (AAS) meeting in science poster sessions.
Participants have primarily been high school teachers
\cite{rebull2018prperpaper1}.  Here, both the words `educator' and
`teacher' are used to refer to the program participants.

Other recent papers describe NITARP in more detail
\cite{rebull2018rtsrenitarp}; see Rebull
\etal~\cite{rebull2018prperpaper1} for more discussion specifically
about the motivations of educators for participating in NITARP, with
ramifications for supporting teachers through a research program.  

In this paper, we ask: How do educator participants describe the major
changes and outcomes in themselves fostered by NITARP?   This study
focuses in particular on the last eight years, specifically the 74
NITARP educator participants from those years. The empirical data are
primarily composed of regular surveys of, and reports from,
participants. Qualitative explorations of the data provide important
knowledge about self-reported teacher participants' learning
experiences in such projects and how the teachers change as a result
of such projects. To emphasize again, this is primarily a summary of
self-report evaluation data, though not all of the themes discussed
here were specifically targeted with survey questions.  Despite the
fact that there is a single project discussed here, in terms of Yin's
terminology \cite{yin2014}, this is a Type IV case study where there
are multiple cohorts, multiple contexts, with multiple units of
analysis.  

Because this research examines how participating teachers describe
their experiences, a social constructivist theoretical framework was
used \cite{koro2009pistemological}.  Constructivism is an
interpretivist theoretical framework; the researchers' goal is to
``describe the practice'' (Koro-Ljungberg
\etal~\cite{koro2009pistemological}, p. 690) within constructivism. To
gain insight as to how teachers’ described changes and outcomes from
participating in NITARP, data were collected from teacher participants
primarily at four waypoints throughout NITARP, but data are also used
from informal and/or smaller-scale surveys of participants and
alumni.  Of the themes described in this paper, about half of them
emerged upon reading all the surveys, and the other half were
specifically probed on the feedback forms. The content validity is
given by triangulation of multiple data sources (surveys at four
waypoints), as well as member checking via participant feedback given
to the researchers \cite{creswell2017qualitative}. 

In this paper, a very brief overview of NITARP is first provided
(Sec.~\ref{sec:overview}); see Rebull
\etal~\cite{rebull2018prperpaper1,rebull2018rtsrenitarp} for more
information. The data are briefly summarized in Sec.~\ref{sec:data}.
Section~\ref{sec:skillsfromnitarp} lists the major changes and
benefits fostered by NITARP, as self-reported by participants,
including a deeper understanding of the nature of science, increased
desire and skill in collaboration, connections to the astronomy
research community, comfort with the unknown, enhanced student
empathy, refinement of their own professional goals, and inclusion of
richer, authentic science activities. Sec.~\ref{sec:summary}
summarizes the results.

\section{NITARP Overview}
\label{sec:overview}

In order to answer the research question about the impact on
educators, at least a simple description of NITARP must be provided. 

NITARP's goal is to provide a long-term PD experience, enabling
teachers to experience the authentic research process. NITARP sets out
to deepen educators' understanding of the nature of scientific
research, and ultimately positively impact their current and future
students via changes in pedagogy.

In this section, the NITARP project is very briefly described. For a
longer description of the program, please see Rebull
\etal~\cite{rebull2018rtsrenitarp,rebull2018prperpaper1}. Because the
present paper includes a discussion on the longer-term impacts of
NITARP on its participants, this section also highlights the ongoing
community of NITARP alumni. Table~\ref{tab:counts} includes some basic
demographic information.

\subsection{Program Context}
\label{sec:context}

The authors of this paper include the NITARP director (LMR) and deputy
(VG), NITARP alumni (WL, DAF), education researchers (TR, WL, MTF,
DAF), staff at IPAC involved in formal and informal education (LMR,
TR, VG, GKS), and professional astronomers (LMR, VG, MTF, GKS).
Because we are so heavily involved in running the program, we can use
the insight provided by our experience to tell a more complete picture
of the NITARP program and what teachers take away from the program. 
Additionally, because the researchers include participants, this
allowed for the researchers to easily clarify participants' response. 
This member checking ensured the validity of these data.  NITARP is
continually changing and adapting to the needs of participants.

Coding was used to analyze the open-ended responses on the surveys
given at the four way-points \cite{creswell2014}.  Participant
responses were obtained, organized, read, coded for initial themes,
then read again, coded for emergent themes. The themes were connected
to paint a more clear picture regarding participants’ views about
their NITARP experience.  

\subsection{High-Level Summary}
\label{sec:highlevelsummary}

NITARP partners a small team of 4-5 educators with a mentor
astronomer. Most ($\sim$70\%) NITARP participants are high school
classroom teachers; there are some middle school classroom educators
and fewer still non-traditional educators. One of the educators on
each team is a mentor teacher who has been through the program before
and acts as a deputy lead of the team. The NITARP year starts with a
trip to the Winter AAS meeting in January.  Right before the AAS
meeting is a day-long ``NITARP Bootcamp'' in which they meet their
team and start to explore both their science and the expectations of
the program.  At the AAS, they see how scientific discourse is
conducted at meetings. They return home and work remotely as a team to
write a proposal; the proposal is reviewed by a panel of educators and
astronomers, and the teams must respond to the comments.  The teams
continue to work remotely through the spring and into the summer. In
the summer, the teams travel to the California Institute of Technology
(Pasadena, CA) with up to four students per teacher for about a week
to work intensively with the data.  They return home and finish the
research through the fall. Finally, they write two posters, one
science and one education, that they take back to the AAS (with
students) in January.  They present their work in science poster
sessions along with other astronomers at the meeting.  NITARP pays all
reasonable travel expenses for the teacher to the first AAS, and the
teacher plus up to two students for the summer visit and the second
AAS.

\subsection{Participant Selection}
\label{sec:participantselection}

NITARP recruits and selects participants from a nation-wide
application process.  Recruitment is primarily through word-of-mouth,
usually via NITARP alumni. Advertising is primarily via email to
astronomy E/PO community contacts, past applicants, and people who
request to be on this mailing list.  There are two waves of email
messaging: once in May, when the call for applications goes out, and
another in August (about 6 weeks before the deadline) when the website
for application submission is opened.  Typically, at least 4 times as
many applications are received as there are available spots.

In terms of participant selection, the ideal participant is someone
who is ready to do research, but has not yet done it. Because the
program is only 13 months long, over which time the participants must
start and complete a research project, the applicant's background is
one of the primary criteria on which applicants are ranked;
participants must be fluent in college-level astronomy.  In an ideal
world, the program would have enough resources to bring any
under-qualified applicants up to speed, but in reality, there are not
enough resources, and participants must learn astronomy via other
opportunities before NITARP. 

The ideal NITARP participant must also not yet have done research. One
of the most important components of the program is the trips to the
AAS at which educators learn how scientific discourse is conducted (at
their first AAS) and present a science poster (at their second AAS).
If an applicant has already attended the AAS and presented a poster,
then they should already know about how scientific discourse is
conducted, particularly when presenting their own work. If an
applicant already has a M.S. or Ph.D. in the physical sciences, they
should already have conducted their own research, and NITARP is
unlikely to be able to teach them very much about how research in
general works. Such overqualified applicants sometimes express a
desire to learn better how to incorporate astronomy data into the
classroom; this particular aspect is not something on which this
program focuses, and so such applicants are referred to other
resources such as the NSTA and AAPT. Some overqualified applicants
could benefit from learning how to access astronomical archives, and
the program refers these applicants to screencapture videos describing
the process.  

Finally, another important criterion for selecting NITARP participants
is their ability to share the experience. Because so few educators can
participate in NITARP each year, each participant must share what they
have learned with others. They need not be currently teaching an
astronomy/physics class nor a student research class to share the
experience with students; for example, math and chemistry teachers
have participated. Teachers can run after-school or weekend astronomy
clubs; teachers may have enough flexibility to include astronomy
within, say, the math or chemistry curriculum.  Applicants who do not
have a venue in which effective sharing of complex concepts to a wide
audience is likely, such as teachers of very young children or home
schooling parents, are unlikely to be selected.

\subsection{NITARP as Effective PD}
\label{sec:nitarpaseffectivepd}

As discussed in Rebull \etal~\cite{rebull2018prperpaper1}, NITARP
aligns well with the characteristics of effective PD. The National
Science Teachers Association (NSTA) Position Statement and Declaration
on Professional Development in Science Education \cite{nstamonograph}
includes the sentence, ``To best serve all students as they learn
science, professional development should engage science educators in
transformative learning experiences that confront deeply held beliefs,
knowledge, and habits of practice.''  Recall the research question:
How do educator participants describe the major changes and outcomes
in themselves fostered by NITARP?   As shown here and below, NITARP
engages educators in transformative learning experiences.

\begin{quote}
{\em The NITARP experience was one of the best professional development experiences I have had. } -- NITARP educator, 2013 class
\end{quote}

\begin{quote}
{\em The NITARP program has opened my eyes to a whole new world —- it has had a enormous impact on what I do, how I do it, and what my students are exposed to.  I really cannot imagine what I would be doing now if I had not gotten involved with this program —- the difference that it has made in my life is truly amazing.} -- NITARP educator, 2016 class
\end{quote}

\begin{quote}
{\em I've been involved in many professional development activities and this is by far the best one I've ever done.} -- NITARP educator, 2010 class
\end{quote}

\subsection{On-going Community}
\label{sec:commofpractice}

A community of practice \cite{lave1991situated} (CoP) is ``a group of
people who share a concern, a set of problems, or a passion about a
topic, and who deepen their knowledge and expertise in this area by
interacting on an ongoing basis'' \cite{wenger2002cultivating}.
Continuing to quote from \cite{wenger2002cultivating}, A CoP's
``purpose is to create, expand, and exchange knowledge, and to develop
individual capabilities'' where members self-select ``based on
expertise or a passion for a topic.'' Holding them together in the CoP
is ``passion, commitment, and identification with the group and its
expertise.'' CoPs ``evolve and end organically as long as there is
relevance to the topic and value and interest in learning together.''
A NITARP CoP is actively supported through regular contact with and
among the alumni.  The NITARP CoP's purpose is to share knowledge
about current events in astronomy (such as press releases or new
journal articles, not ``what's in the sky this month,'' which can be
found via many other venues), in astronomy education, and other items
such as new opportunities to participate in other programs, as well as
events within the NITARP year.  The program maintains a mailing list
where opportunities are shared and where teachers can ask for help. 
The members are selected to join via selection for NITARP, but they
are free to unsubscribe from the mailing list; nearly all of the
alumni have maintained their subscription to the mailing list, even
through job changes (or retirement), as manifested by changes in email
addresses in the list. Thus, the NITARP CoP meets the criteria of
self-selection based on passion and commitment; the group and its
expertise is seen as valuable. Mail traffic on this list is largely
from the NITARP management to the community, but at least once every
4-8 weeks, queries from alumni go out asking for help in solving a
problem, say, in the classroom or getting students more involved in
research, etc. Additional teacher-to-teacher direct contact occurs, as
manifested by subsequent collaboration and products. There is no
restriction on topics in the list, and the topics covered in the list
evolve with time, meeting the last quality listed for a CoP.

Further, Ref.~\cite{wengertraynerweb} describe CoPs as ``Communities
of practice are groups of people who share a concern or a passion for
something they do and learn how to do it better as they interact
regularly.'' They go on to list three characteristics: (1) ``The
domain: [...] It has an identity defined by a shared domain of
interest. Membership therefore implies a commitment to the domain, and
therefore a shared competence that distinguishes members from other
people.'' (2) ``The community: In pursuing their interest in their
domain, members engage in joint activities and discussions, help each
other, and share information. They build relationships that enable
them to learn from each other; they care about their standing with
each other.'' (3) ``The practice: [...] Members of a CoP are
practitioners. They develop a shared repertoire of resources:
experiences, stories, tools, ways of addressing recurring problems --
in short a shared practice.'' NITARP combines all three of the
defining characteristics for a CoP. Members of the CoP clearly shares
a passion for astronomy teaching, and they learn from each other how
to improve their teaching and astronomy research skills by interacting
regularly. They have a shared competence that distinguishes them from
other educators; many NITARP educators win awards (see the incomplete
list on the NITARP website). They engage in joint activities and help
each other, sharing information. They share experiences, tools, and
approaches for doing research with students. More focused subgroups
appear within the community, such as those in elite private schools or
those in remote rural public schools, to address those groups'
specific needs.  NITARP meets the qualities listed for a CoP.

In recent years, a `continuing education' video series has been
initiated for the NITARP CoP, sharing new tools and data releases. 
These videos are posted publicly via YouTube.  Many videos have taken
on a life of their own; the videos on ds9 (a tool for viewing FITS
\cite{1981A&AS...44..363W} images) were posted -- by the ds9 staff --
on Harvard's main ds9 page. Additionally, through affiliation with
NASA-JPL, the alumni community also has access to telecons aimed at a
broader audience and covering current events in space exploration and
astronomy.   

As discussed in Sec.~\ref{sec:researchcommunity} below, many of the
alumni continue to do research or similar activities, either with
their original team or with a new team composed of other alumni. This
ongoing involvement in research activities also contributes to the
CoP. 

\begin{quote}
{\em Not only can I analyze astronomical data to find scientifically useful results, but I can publish my work as a poster and be part of this community. I did not feel like a stranger or usurper or even out of place –- it felt like I belonged.} -- NITARP educator, 2014 class
\end{quote}

\begin{quote}
{\em The reality of taking the children to the AAS was more than I could have imagined. The excitement on their faces as they saw the community that they were now a part of and the fact that they were able to experience the independence of presenting the work to professionals, and feel successful doing so. I watched students who came in apprehensive and nervous and they blossomed as they got comfortable and practice. It was beautiful to see.} -- NITARP educator, 2016 class
\end{quote}

\begin{quote}
{\em  I think that future support will be needed if I am able to try independent work –- but that might be from the great network of alumni and astronomy community as well.} -- NITARP educator, 2017 class
\end{quote}

\begin{quote}
{\em I am very grateful to be again part of such an amazing and dynamic group of people and always love how supportive and encouraging the NITARP community is.} -- NITARP mentor educator, 2017 class
\end{quote}

\section{Data}
\label{sec:data}

The data that are the primary focus of this analysis are discussed in
detail in Rebull \etal~\cite{rebull2018prperpaper1}; here we briefly
summarize the data. Important numbers are summarized in
Table~\ref{tab:counts}. Themes discussed in the rest of the paper are
summarized in Table~\ref{tab:themes}.

Since 2005, 103 teachers have participated in NITARP (or its immediate
predecessor). We use detailed, written survey data from 74 teachers
collected over the most recent 8 years, 2010-2017.  At up to four
waypoints during each NITARP year, surveys were collected from
participants :
\begin{itemize}
\item {\bf Pre-AAS}: Before their first AAS (initiated with the 2015 class);
\item {\bf Post-first-AAS}: After the NITARP Bootcamp and their first AAS;
\item {\bf Summer}: After the summer work session (includes teachers and students who participate in this visit);
\item {\bf Post-second-AAS}: After their second AAS at which they presented their results (includes teachers and students).
\end{itemize}

Nearly twice as many educators were involved in 2010-2013 as compared
to 2014-2017 \cite{rebull2018prperpaper1}; see summary in
Table~\ref{tab:counts}. The surveys were substantially changed in the
middle of 2014 as a result of `boots-on-the-ground' experiences in the
first four years, coupled with a better understanding of the education
research literature (see Rebull \etal~\cite{rebull2018prperpaper1} for
a complete list of survey questions). Despite 95-100\% participation
from teachers on the surveys in general, data from the first four
years (2010-2013, but particularly 2010) are less complete than data
from the most recent four years (2014-2017). Therefore, by number,
there are more people in the earlier years, but more and better
surveys (and answers) in the later years. Of the 74 educators, 70\%
are (or were at the time) high school educators, 65\% are/were public
school educators, and 57\% are women \cite{rebull2018prperpaper1}.

While the educators originate from a wide variety of states and types
of schools/programs, it is frequently the case that the participants
are already quite accomplished educators, and they tend to seek out
opportunities to learn and improve their practice. In the context of
the present work, which is heavily based on self-reported changes, it
is important to note that these teachers are highly capable of
recognizing changes in their teaching approach and/or philosophy.
These individuals' ability to reflect upon their growth, critically
analyze themselves, and answer with great richness and detail, lends
validity and weight to their self-reported changes.

All answers to the collected surveys were examined, and emergent
themes were sought. Themes of broad interest (``What is `real
astronomy'?'') were probed with explicit questions (survey questions
appear in their entirety in Rebull
\etal~\cite{rebull2018prperpaper1}). In the context of this work, new
themes emerged upon reading the responses in aggregate, as described
below. The surveys were then re-read to look for those emergent
themes. In some cases, the answers were iteratively coded for emergent
themes; also see Rebull \etal\ \cite{rebull2018prperpaper1}. Themes
are listed in Table~\ref{tab:themes} with an indication as to whether
they were identified {\em a priori} or emerged as part of this work.

Quotes used in this paper primarily come from these surveys. Some
quotes are extracted from email from teachers to management or to
mentor scientists. There was also a brief survey that attempted to
quantify NITARP impact from 2005-2013; results from that appear in
Rebull \etal~\cite{2014AAS...22324901R}, available on the NITARP
website. Some quotes from that survey appear in this paper. Finally, a
few quotes come from informal surveys that some teams did as part of
their education AAS posters. All of these education posters are
available on the NITARP website. The fact that similar themes can be
identified in quotes originating in very different surveys lends
support to these results.

\begin{table*}
\caption{Counts\label{tab:counts}}
\begin{ruledtabular}
\begin{tabular}{cccp{12cm}}
\# Counted & Explicit  & Calculated  & Notes\\
 &  Fraction &  Fraction  & \\
\hline
\multicolumn{4}{l}{Demographics (Sec.~\ref{sec:data} and \cite{rebull2018prperpaper1})}\\
103 & \ldots & \ldots   & Unique participants 2005-2017\\
74 & \ldots & \ldots &  Unique participants 2010-2017 with surveys at up to 4 waypoints\\
51 & \ldots & \ldots & Unique participants 2010-2013 with surveys at up to 3 waypoints\\
27 & \ldots & \ldots &  Unique participants 2014-2017 with surveys (better questions; added pre-AAS) at up to 4 waypoints; includes mentor teachers selected out of prior 4 years, so 51+27 != 74\\
23 & \ldots & \ldots &  Unique participants 2014-2017 with surveys; counts mentor teachers only once, so 51+23 = 74\\
52 & 52/74 & 0.70 & HS educators \\
48 & 48/74 & 0.65 & public school educators \\
42 & 42/74 & 0.57 & women educators\\
\hline
\multicolumn{4}{l}{Nature of Science (Sec.~\ref{sec:natureofscience})}\\
9	&	9/74	&	0.12	&	no information, counting over all\\
10	&	10/74	&	0.14	&	no change, counting over all\\
35	&	35/74	&	0.47	&	some change, counting over all\\
20	&	20/74	&	0.27	&	major change, counting over all\\
9	&	9/51	&	0.18	&	no information, counting over first 4 NITARP years\\
7	&	7/51	&	0.14	&	no change, counting over first 4 NITARP years\\
22	&	22/51	&	0.43	&	some change, counting over first 4 NITARP years\\
13	&	13/51	&	0.25	&	major change, counting over first 4 NITARP years\\
0	&	0	&	0	&	no information, counting over second 4 NITARP years\\
3	&	3/23	&	0.13	&	no change, counting over second 4 NITARP years\\
13	&	13/23	&	0.57	&	some change, counting over second 4 NITARP years\\
7	&	7/23	&	0.30	&	major change, counting over second 4 NITARP years\\
\hline
\multicolumn{4}{l}{Qualities of an astronomer (Sec.~\ref{sec:qualitiesofastro})}\\
15	&	15/31	&	0.48	&	Number of surveys out of 31 available surveys with this explicit question mention teamwork, collaboration, etc.\\
19  &  19/31   & 0.61 & Number of surveys mentioning patience, persistence, etc.\\
12 & 12/31 & 0.39 & Number of surveys mentioning creativity, etc. \\
\hline
\multicolumn{4}{l}{Best thing about the trips (Sec.~\ref{sec:bestthingtrips})}\\
28	&	28/74	&	0.38	&	Number of educators listing `working with students'\footnote{The fraction is out of all educators, as opposed to all encoded responses; some educators list more than one thing as the best thing about the trip, so the sum of instances here is $>$74, but the fraction is the fraction of educators citing this item, as opposed to the fraction of encoded responses. This also means that the total of the left hand column for this subsection is $>$74.} \\
14	&	14/74	&	0.19	&	Number of educators listing `doing real research'\\
6	&	6/74	&	0.08	&	Number of educators listing `learning new things'\\
11	&	11/74	&	0.15	&	Number of educators listing `doing the hands-on work'\\
36	&	36/74	&	0.49	&	Number of educators listing `working as a team' or `teamwork' \\
5	&	5/74	&	0.07	&	Number of educators listing `meeting new people' \\
9	&	9/74	&	0.12	&	Number of educators listing `meeting and/or working with scientists' \\
15	&	15/74	&	0.20	&	Number of educators listing `taking the tour(s)' \\
9	&	9/74	&	0.12	&	Number of educators listing `balance of work and play' \\
\hline
\multicolumn{4}{l}{Comfort with the Unknown (Sec.~\ref{sec:unknown})}\\
27	&	27/64	&	0.42	& Lower limit; did not ask a question that routinely resulted in this kind of answer\\
\hline
\multicolumn{4}{l}{Student Empathy (Sec.~\ref{sec:studentempathy})}\\	
5	&	5/74	&	0.07	&	Lower limit; did not ask a question that routinely resulted in this kind of answer\\
\hline
\multicolumn{4}{l}{Professional Growth (Sec.~\ref{sec:jobs})}\\
61	&	61/74	&	0.82	&	Looking for opportunities to learn/grow/change (probably lower limit, because asking only recently)\\
9	&	9/74	&	0.12	&	Lower limit; NITARP had significant role in career change\\
5	&	5/74	&	0.07	&	Lower limit; NITARP prompted/encouraged going to graduate school in education\\
10	&	10/74	&	0.14	&	Lower limit; used the words ``life changing" to describe their NITARP experience\\
\hline
\multicolumn{4}{l}{Getting better science in classroom and PD (Sec.~\ref{sec:betterscience})}\\
44	&	44/74	&	0.59	&	Probably lower limit; expressed desire to get into classroom, as opposed to witnessed in classroom\\
\end{tabular}
\end{ruledtabular}
\end{table*}

\begin{table*}
\caption{Themes\label{tab:themes}}
\begin{ruledtabular}
\begin{tabular}{lcccp{12cm}}
Theme & Section  & \# quotes\footnote{Number of quotes included in the paper, not number of total quotes available.} & {\em a priori} or emergent & Notes\\
\hline
(NITARP as effective PD) & \ref{sec:nitarpaseffectivepd} & 3 & \ldots & (from NITARP summary)\\
(NITARP and ongoing community) & \ref{sec:commofpractice} & 4 & \ldots & (from NITARP summary)\\
Nature of Science & \ref{sec:natureofscience} & 12 & {\em a priori}\\
Qualities of an Astronomer & \ref{sec:qualitiesofastro} & 13 & {\em a priori} & emergent first 4 yrs; {\em a priori} second 4 yrs\\
Collaboration and Sharing & \ref{sec:sharing} & 8 & emergent& \\
Links to Astronomy Research Community & \ref{sec:researchcommunity} & 4 & emergent&   \\
`Best Thing about the Trips' & \ref{sec:bestthingtrips} & 6 & {\em a priori}& \\
Comfort with the Unknown & \ref{sec:unknown} & 12 & emergent& \\
Student Empathy & \ref{sec:studentempathy} & 2 & emergent&  \\
Professional Growth & \ref{sec:jobs} & 4 & {\em a priori} & emergent first 4 yrs; {\em a priori} second 4 yrs\\
Better Science in their Classrooms and PD & \ref{sec:betterscience} & 8 & {\em a priori} & emergent first 4 yrs; {\em a priori} second 4 yrs\\
\end{tabular}
\end{ruledtabular}
\end{table*}


\section{Findings: Major Changes and Benefits Fostered by NITARP}
\label{sec:skillsfromnitarp}

This paper has a primary goal of probing how educator participants
describe the major changes and outcomes in themselves fostered by
NITARP. 

In looking at the survey results and thinking about the higher-level
and longer-term gains seen in the educator participants, there are
some noticeable trends. The skills that are fostered in these
educators, or skills that the program hopes to foster, are also to
some extent characteristics of successful professional astronomers. In
the application process, teachers who are ready to learn these skills,
or have already started to learn these skills, are actively sought.
There is evidence from the feedback forms that they learn these skills
over the year, but there is also evidence of continued development of
these skills among the alumni community, via emails to the NITARP
mailing list and continued communication with the scientists. 

This section discusses major changes and benefits fostered by NITARP
grouped under four major categories: perceptions of science and
scientists; impact of collaborations; views of learners; pedagogical
changes and professional growth.

\subsection{Perceptions of Science and Scientists}

\subsubsection{Nature of Science}
\label{sec:natureofscience}

Understanding the nature of science is often cited as a primary goal
of many different educational programs aimed at teachers and/or
students. The nature of science can be interpreted in many ways
\cite{lederman1992students,lederman2007nature}, but here we adopt the
meaning as in Lederman \cite{lederman2007nature}, where `nature of
science' means how science accumulates evidence and science as a way
of learning about the world.  

Because the educators have to start and finish a research project in a
year, participants are selected to already be fluent in college-level
astronomy. Therefore, many of them already are confident before
starting the program that they know about astronomers (``astronomers
are real people''; see Sec.~\ref{sec:qualitiesofastro} for more on
this) and astronomy (and how science works in general). Since one of
the major program goals was to change how teachers think about science
and scientists, questions probing this were included on the feedback
forms starting early in the program. Survey questions appear in their
entirety in Rebull \etal~\cite{rebull2018prperpaper1}, but the ones
most relevant to this discussion are, ``Did this experience change the
way you thought about astronomy or astronomers?'' and to a lesser
extent, ``Did you do anything on this visit that you expected would be
part of scientific research? Or anything that you did not think would
be part of scientific research?''

Based on the open-ended responses in their surveys, the educators can
be placed into one of four bins: 
\begin{enumerate}
\item ``No information'': Those educators for whom there is no information, where literally this question was not asked or no answer to that question was provided.
\item ``No change'': Those educators reporting no change in understanding. For example, in response to ``Did this experience change the way you thought about astronomy or astronomers?'' an educator might respond, simply, ``No.''  (Also see quotes below.)
\item ``Some change'': Those educators reporting some change in understanding, or a more nuanced understanding.  For example, in response to ``Did this experience change the way you thought about astronomy or astronomers?'' an educator might respond, simply, ``Not really, but I had no idea that there were so much public data available,'' or ``Not really, but I didn't know how much computer programming was involved.'' (Also see quotes below.)
\item ``Major change'': Those educators reporting a major change in understanding. For example, in response to ``Did this experience change the way you thought about astronomy or astronomers?'' an educator might respond, ``Yes, it has revolutionized my understanding!'' (Also see quotes below.)
\end{enumerate}

\begin{figure}
\includegraphics[width=3in]{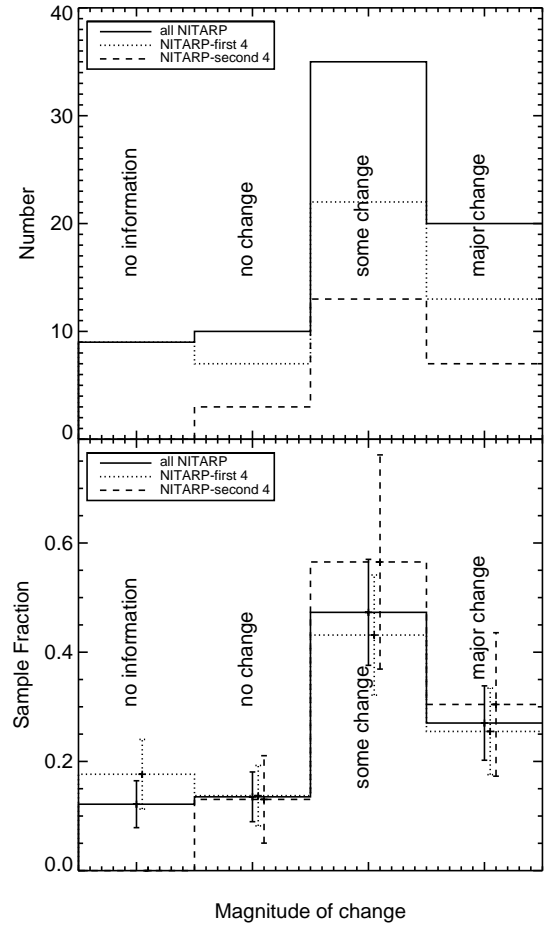}
\caption{Numbers (top) and fractions (bottom) of educators who report changes in their understanding of the nature of science. The first bin is those educators for whom there is no information, the second bin is those reporting no change in understanding, the third bin is those reporting some change in understanding (a more nuanced understanding), and those who report a major change in understanding. Because the wording of our questions changed, plotted here are the aggregate counts (solid line), but also the first four years of NITARP (``NITARP-first 4'', dotted line), as distinct from the second four years of NITARP (``NITARP-second 4'', dashed line). Errors shown in the second panel assume Poisson statistics, and are slightly offset for clarity. About 74\% ($\pm$13\%) of our educators report either some or a major change in their understanding of the nature of science.
\label{fig:natureofsci}}
\end{figure}

Overall, for 12\% of the educators, there is no information about this
topic, either because they were not asked, or they didn't answer the
question; all of these educators are from the first four years.  About
14\% ($\pm$5\%, assuming Poisson statistics) of educators report that
this program did not change their opinion of science/scientists or
astronomy/astronomers; they already felt that they knew what real
science/scientists were like, and their experience did not change
that. Just under half (47\% $\pm$ 10\%) report that there was some
change in their opinion; for example, they report an increased
appreciation for how much data are in astronomical archives, or that
there was a lot more programming than they expected, etc.  Finally,
27\% ($\pm$7\%) report major changes in their understanding, a
revolution in their thinking.  Fig.~\ref{fig:natureofsci} visualizes
the fractions of the samples in each bin. Because the wording of our
questions changed somewhat, plotted here are the aggregate counts
(solid line), but also the first period of four years (``NITARP-first
4'', dotted line), as distinct from the second period of four years
(``NITARP-second 4'', dashed line).

\begin{quote}
{\em This experience has completely changed my once shallow view of astronomy and astronomers.} -- NITARP educator, 2015 class (major change)
\end{quote}

\begin{quote}
{\em Before ever having experienced an American Astronomical Society meeting I thought I was well versed in the astronomer's culture. [...] My experience has been one of culture shock. Astronomers, all of whom are scientists, can be personal, funny, and outright social beings. The nature of their work - retracing their steps for accuracy, being critical of fellow colleagues, and looking to develop the next best project that has not been accomplished already - requires astronomers to discuss, inquire, and exchange their ideas with one another.} -- NITARP educator, 2013 class (major change)
\end{quote}

\begin{quote}
{\em One evening, while working on some homework, I had the realization that THIS WAS REAL. There is no right answer, in fact, no one knows the answer. I can't just go and ask someone the answer. It was like a light bulb went off and I experienced a feeling of excitement and also felt a little bit scared. I thought to myself – Is this how astronomers feel about their work? It was a great feeling and exciting that I too am part of this now.  } -- NITARP educator, 2012 class (major change)
\end{quote}

\begin{quote}
{\em The entire experience was ``real astronomy.''  Nothing was canned.  None of us in the room knew what the ``final answer'' was.  Students really buy into the fact that this is real research.  We may only find five new young stars, but when we do, we will be the only people on the planet that know that they are there.  How cool is that?  They get to go through the process of science and learn, as Feynman would put it,``the kick in the discovery!''} -- NITARP educator, 2012 class (major change)
\end{quote}

\begin{quote}
{\em  When my past students did astronomy research projects, they used data that they themselves collected[...]  After working at Caltech with the group, though,  I have come to realize that what my students have been doing previously were small projects compared to our [NITARP] study—-they were really just glorified lab activities [because I knew what they were going to find before they started].  I have been giving a lot of thought to this since I returned home and am planning major changes in the sort of projects that my research students will be working on in the future. } -- NITARP educator, 2013 class (major change)
\end{quote}

\begin{quote}
{\em Well, here's the thing: as an engineer familiar with rigorous mathematical modeling and iterative problem solving, I thought I could do science; I thought was basically the same thing, only with theories instead of problems. Thinking I ``had what it takes to be a scientist'' turns out to have been hubris. In other words, I would have answered this question with [I know what science is] and I would have been wrong.} -- NITARP educator, 2016 class (major change)
\end{quote}

\begin{quote}
{\em  The process of gathering and analyzing data was very important to help show my kids what real research is like.  And since our data didn't come out nice and neat like some labs do, it really helped push my kids to think outside the box.} -- NITARP educator, 2013 class (some change; note this person is focused on student learning in this quote)
\end{quote}

\begin{quote}
{\em I never realized how much computer programming is done in Astronomy. I think this will help me reach out to students who might not be interested in `science.' These students may not realize that their programming skills are vital for analyzing astronomical data.} -- NITARP educator, 2011 class (some change)
\end{quote}

\begin{quote}
{\em It has changed a little-- I now realize that the data does not have to actually be collected by the scientist, but can be collected by anyone.  In fact, much science is now done by ``data mining" where the data may have already been collected, often for some other purpose, but can be mined for things that the original project did not conceive of.} -- NITARP educator, 2016 class (some change)
\end{quote}

\begin{quote}
{\em I have always loved astronomy and have had great interactions with many great people in the field so my thoughts on [astronomers] are just as positive as ever.} -- NITARP educator, 2017 class (no change)
\end{quote}

\begin{quote}
{\em No [it did not change my thoughts on astronomy or astronomers].  I'm just really impressed with how much everybody loves what they do.} -- NITARP educator, 2011 class (no change)
\end{quote}

\begin{quote}
{\em I have been around astronomers and astronomy enough to only have my current views reinforced, which are that as a discipline, astronomy has some of the kindest and most passionate scientists around. They are great fun to work with.} -- NITARP educator, 2016 class (no change)
\end{quote}

In summary, then, three-quarters (74\%$\pm$13\%, assuming Poisson statistics) of the educators report either major or some change in their understanding of the nature of science as a result of their experience. There is some indication that the changes in the participating students' understanding of the nature of science may be substantially more profound, and we leave further discussion of this to a future paper. Similarly, changing teachers' understanding of science is most likely distinct from making sure that this understanding is conveyed to subsequent students in the classroom (those who didn't come on the NITARP trips) or other educators; this is also beyond the scope of the present work. Buxner \cite{buxner2014exploring} found that strong changes in teachers' understanding of science inquiry did not commonly occur in three different summer teacher research programs, so it is important, but beyond the scope of the present work, to follow up on the longer-term ramifications of the NITARP educators' self-reported changes in understanding, specifically as it applies to the classroom. 

\subsubsection{Qualities of an Astronomer}
\label{sec:qualitiesofastro}

As discussed in Sec.~\ref{sec:participantselection}, the participants
are selected to have a college-level understanding of astronomy, and
come into the program often confident that they know all about
astronomy and astronomers. However, in many cases, NITARP up-ends that
understanding. At the beginning of NITARP (and in the preceding
years), it had been assumed that the educators already knew that
`astronomers are real people,' and that astronomers (at least for the
most part) do not look like the popular scientist stereotype of an
older white man with a lab coat and bad hair (e.g.,
\cite{chambers83,yang2017exploring}).   However, in the first four
years of feedback forms, it became apparent that some educators did
not know this, and so in the most recent four years of data, the
surveys explicitly asked, ``What qualities do you think are important
to be an astronomer?''  After careful review of participants’
responses, additional themes emerged. For example, see the second
quote in Sec.~\ref{sec:natureofscience} above; it says, ``Astronomers,
all of whom are scientists, can be personal, funny, and outright
social beings.'' In answers to the questions about things participants
thought were surprising, or things that they noted particularly after
their first AAS trip, it can be seen that many participants had
misconceptions about what astronomers look like or how they interact.
NITARP has dispelled these misconceptions.

\begin{quote}
{\em The route into astronomy is more varied than I thought. More people have access than I anticipated. I would never have thought someone who was older, coming from community college would end up with a Ph.D.\ in astronomy, working at Caltech. Or that a girl who thought she couldn't do math well into college would end up in astronomy. These stories were encouraging.} -- NITARP educator, 2012 class
\end{quote}

\begin{quote}
{\em I am used to seeing older people as astronomers as we watch NOVA and other videos or read about past astronomers in class.} -- NITARP educator, 2012 class
\end{quote}

\begin{quote}
{\em Astronomers are a remarkably collaborative lot. I knew this, but I was amazed by how friendly everyone was. For the most part I was clear that I was a teacher, and they probably had nothing to gain by talking to me. For most people, this was not a deterrent.} -- NITARP educator, 2012 class
\end{quote}

\begin{quote}
{\em I learned that astronomers are much more down-to-earth than I envisioned a lot of them to be and so many of them are very willing to go out of their way to explain things when we have questions.} -- NITARP educator, 2011 class
\end{quote}

\begin{quote}
{\em This experience certainly has changed my thoughts about astronomy and astronomers. I really did not know what they did except teach college classes.  I enjoyed seeing the less formal and family friendly atmosphere at [Caltech].  This experience will be shared with my students for some time. } -- NITARP educator, 2015 class
\end{quote}

\begin{figure}
\includegraphics[width=3.4in]{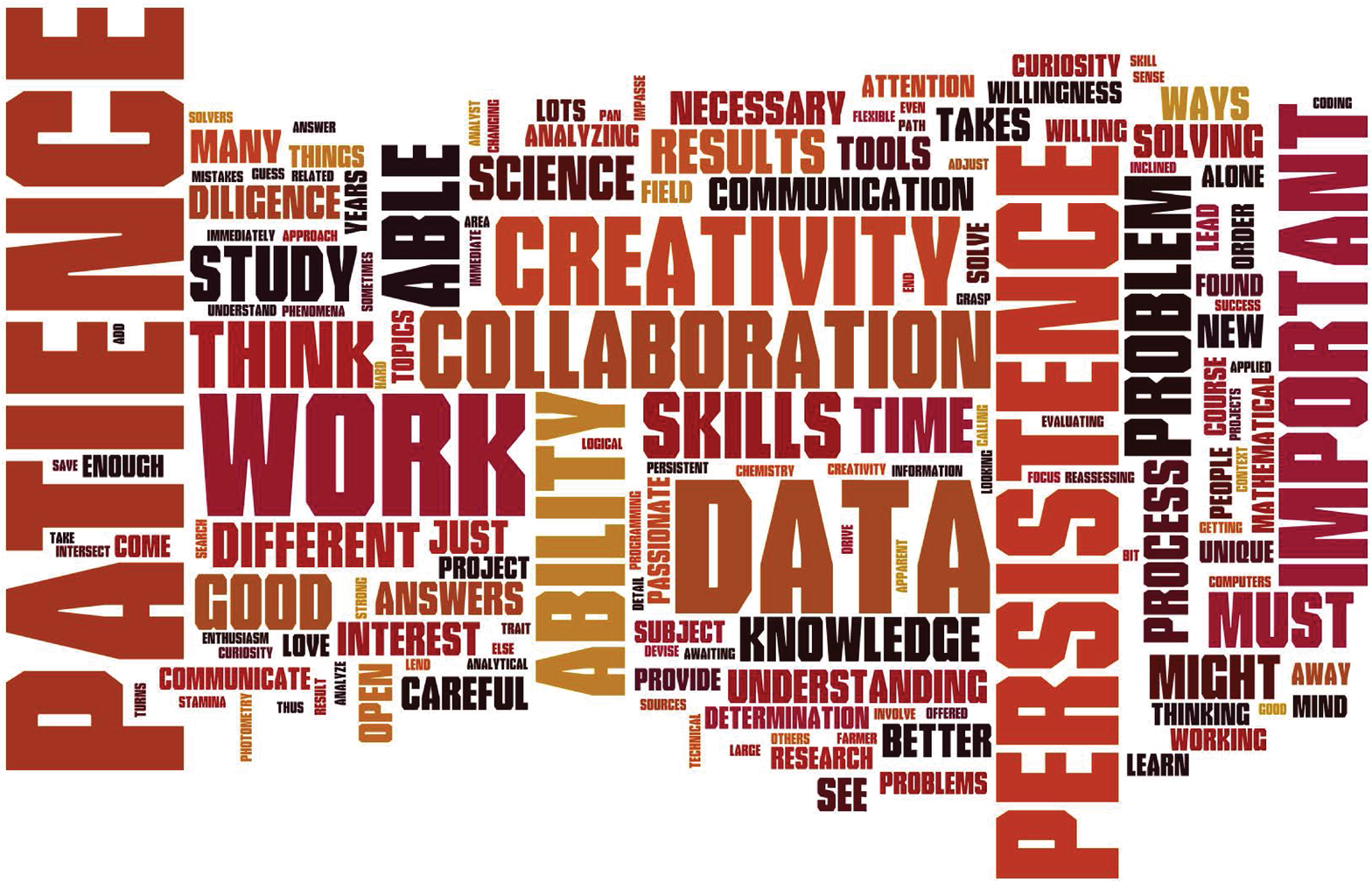}
\caption{Qualities of an astronomer. Words that have prominence in this word cloud include patience, persistence, creativity, and collaboration. 
\label{fig:astronomerqualities}}
\end{figure}

More broadly, Figure~\ref{fig:astronomerqualities} is a word cloud of
the words used in responses from 2014-2017 to this question ``What
qualities do you think are important to be an astronomer?''  Some of
the words that appear there prominently are ones that people might
likely use to describe scientists of any sort -- work, data,
important, problem, study, think, ability/able, skills, and time. A
Google search reveals that these kinds of words come up frequently
when referring to scientists. However, words that might not come to
mind for the average person (or appear more rarely in a Google search)
but that have prominence in this word cloud include patience,
persistence, creativity, and collaboration. 

Specifically in the context of Section~\ref{sec:sharing} below on
collaboration, since collaboration and sharing is often new to these
educators, it can be seen that over the years for which there are
complete surveys, 30\% of the educators that filled out a survey
explicitly asking about the qualities of an astronomer list
collaboration as important.

\begin{quote}
{\em Astronomers need to be independently motivated (be able to work alone with a strong drive) and to also work well in teams (and not just with other scientists).  They are problem solvers and love a good mystery.  Also, they need to have the patience to stick with a project for years.  And they should be able to juggle more than one project at once.} -- NITARP educator, 2016 class
\end{quote}

\begin{quote}
{\em Real astronomy involves being able to problem solve and think critically, apply process skills, and communicate effectively.  There has to be passion about the subject matter as well as a dedication and certain element of dedication to the field. Collaboration is huge...being able to work well with other people. Lots of computer time!} -- NITARP educator, 2010 class
\end{quote}

\begin{quote}
{\em Based on my experience so far in this program, I have found that “real astronomy” involves a lot of data processing. In addition, I have found that collaboration is a very important aspect of the process.} -- NITARP educator, 2013 class
\end{quote}

Similarly out of that same set of feedback forms with the explicit
question, 61\% of the educators mentioned patience, persistence, or
other similar qualities as important qualities of an astronomer.
Finally, 39\% mention creativity. 

\begin{quote}
{\em There are lots of qualities that are important to an astronomer, but two that come to mind are persistence and diligence.  Sometimes, the apparent path to solving a problem turns out to lead somewhere else (or not lead anywhere at all).  The astronomer has to be aware of this and know when to change course and try another approach -- sometimes, this has to be done over and over again before the research problem starts to show results.} -- NITARP educator, 2016 team
\end{quote}

\begin{quote}
{\em  Real astronomy is done in data analysis. This is not what the public sees or imagines. I think that this is one of the great values of the program, particularly for students, but for teachers as well. Participating in research projects is always eye-opening and exciting, but it involves a great deal of hard work and creativity. This last aspect, creativity, is an area where scientists typically receive little credit, but it is where they truly excel. Science, including astronomy, seems like a stodgy and non-creative endeavor on the outside, particularly with the focus on STEM as separate from the arts, which are considered creative.} -- NITARP educator, 2016 class
\end{quote}

\begin{quote}
{\em I think more teachers and people in general should be exposed to the process [of research] more so they can understand how dynamic it can be, how much creativity and tenacity is needed, and how NOT like traditional text books it is.} -- NITARP educator, 2017 class
\end{quote}

\begin{quote}
{\em Astronomy research often involves using public astronomy archives, and sometimes it only involves using archives.  Success in astronomy is not just a result of brilliance (though brilliance doesn't hurt).  Success also requires quite a bit of persistence.} -- NITARP educator, 2015 class
\end{quote}

\begin{quote}
{\em Astronomers need to be curious and persistent.  Persistence might involve anything from seeking funding sources to awaiting new technologies that could provide the data you need.} -- NITARP educator, 2015 class
\end{quote}

Specifically because the teachers largely come into NITARP believing
themselves already knowledgeable about the nature of astronomer and
astronomers, there has been an expectation that the impact on the
participating students' understanding of astronomers was more
substantial. Schneider \cite{schneider2017} explored the impact of
students' understanding after their teachers went through an RET
program, and found substantial changes on students, though no impact
on the teachers, perhaps due to sample size. It is likely that those
RET teachers similarly believed themselves already knowledgeable, and
there may not have been enough educators to find the ones that had
their perceptions overturned.  Indeed, as in the section above on the
nature of astronomy, there is some indication that the changes in the
participating students' understanding of what astronomers look like
(and do) may be substantially more profound. In one case, a teacher,
having read his student's surveys before turning them into NITARP,
noted that evidently he had done a poor job in conveying to his
students what astronomers look like, because he knew and they didn't.
We leave further discussion of the impact on students to a future
paper.

\subsection{Impact of Collaborations}

\subsubsection{Collaboration and Sharing Itself}
\label{sec:sharing}

For many astronomers, collaboration with colleagues is an integral
part of being a scientist. Few astronomers publish single-author
papers anymore; most papers are written by collaborations of people 
\cite{frogel2010astronomy}. Indeed, studies suggest that the more
diverse a research team is, the more significant the results (or at
least the more citations the paper gets); see  Abt
\cite{abt2017citations} for a discussion of this within astronomy, or
Freeman \& Huang  \cite{freeman2014strength} for a more general
assessment.

In contrast, many science teachers work in isolation, though research
has shown that collaboration is important (e.g.,
\cite{lave1991situated,slavit2009perspectives,nelson2005knowledge}).
NITARP is, by its very nature, collaborative; the teachers work in
teams. Even if they approach the project initially thinking that they
will work on their own within the team, most teachers find rather
rapidly that the teamwork itself is a powerful motivator
\cite{rebull2018prperpaper1}.  More specifically, these teams require
distance collaboration, which is even more rare among teachers.
Participants learn how to take advantage of distance collaboration
tools to expand their own network, within and beyond the program.

The importance of working in a team in NITARP, as well as the
importance of collaborating and sharing process, progress, and
results, was not at all understood prior to this analysis. It emerged,
powerfully and obviously, upon reading all the survey answers. 

\begin{quote}
{\em Real astronomy involves working as a team to find an answer.} -- NITARP educator, 2011 class
\end{quote}

\begin{quote}
{\em I love the connections and sharing that happen between teachers, students, and mentors.  This is the part that I believe makes a successful PD or E/PO project very successful -- the interactions and networks that are formed between participants at all levels.  It builds a working comfort level and common ground that helps drive and support further collaboration and communication.} -- NITARP educator, 2017 class
\end{quote}

\begin{quote}
{\em NITARP brings science directly into the hands of teachers and students, and demonstrates how science is an active, collaborative, and evolving effort.} -- NITARP educator, 2014 class
\end{quote}

\begin{quote}
{\em My students saw a community of people who truly love what they do and are willing to explain it. They made many contacts and saw what science is all about -- sharing discoveries and collaboration.  Most importantly, they discovered that they can do this themselves, that they can belong to this community as well. } -- NITARP educator, 2011 class
\end{quote}

Particularly early on, it was not anticipated that the team itself
(and the CoP composed of alumni) provides a very useful support
network, not just for NITARP activities, but other activities relating
to education.  Some teachers are surprised at how quickly they `bond'
with their team, and many continue to collaborate (sympathize,
bolster, share resources, etc.) after their research experience year;
some educators forge friendships/collaborations with alumni from their
own class as well as other classes. 

\begin{quote}
{\em [Another teacher] and I actually shared curriculum for astronomy and bounced ideas off one another.  The collaborating is priceless. } -- NITARP educator, 2014 class
\end{quote}

\begin{quote}
{\em The best thing about the trip was the chance to interact with others who are trying to do the same things that I am trying to do.  No one else around me tries to do student research (even though I have tried to get other teachers involved), not in my district nor in any of the surrounding ones.   It was great to spend time with other teachers (and their students) who are trying to accomplish the same things that I am trying to do.} -- NITARP educator, 2013 class
\end{quote}

\begin{quote}
{\em I really enjoyed working with the [other] teachers. It was important to help each other out and realize that you are going to make mistakes and hopefully one of the other teachers can bring you up to speed on the different areas.} -- NITARP educator, 2011 class
\end{quote}

Even those teachers who were on teams that `broke' (for whatever
reasons) seem to still benefit from the network and support structure
provided. The CoP is strong enough to maintain the support network
even when the team (nominally an educator's closest collaborators)
struggles.

\begin{quote}
{\em Personally, all the professional contacts (at AAS meetings and
other teacher NITARP participants) plus access to other programs via
these contacts and your email blasts have been phenomenal and expanded
what I have been able to accomplish. Through these contacts and
programs, I have been able to bring other astronomy related programs
to students here at the high school and I have gained new knowledge,
too. Our students, over the past few years since I participated in
NITARP, have directly benefited from my being accepted into the
program. You are a great asset and I appreciate all that you do!} --
NITARP educator, on a 2013 team with challenges, from an email in 2016
\end{quote}

NITARP participants have volunteered the insight that sharing and/or
collaboration among the high school science teachers in their school
or district is far from widespread, and that NITARP really
demonstrates in concrete fashion the power of collaboration.

\subsubsection{Links to Astronomy Research Community}
\label{sec:researchcommunity}

Alumni seek the intellectual stimulation of the NITARP and wider
astronomy communities. Many alumni come back to the AAS, raising their
own money to attend. There have not been careful records about how
many self-funded alumni return; it was first noticed that there was a
significant increase in the number of alumni at the 2014 AAS meeting. 
At the 2015, 2016, and 2017 January meetings, for which there are at
least partial records, the number of self-funded alumni educators was
comparable to each class size; that is, there are 8-9 teachers per
class, and about that many alumni, at least, who have paid their own
way back to the AAS. So, out of the NITARP-affiliated educators at the
meeting, about a third are alumni. (Because teachers in the class that
is finishing up bring many students, and because alumni teachers often
bring even more students than they did in their research experience
year, there are always more NITARP-affiliated students than
educators.) Some alumni come only to AAS meetings near them (for
financial reasons); others come to every winter AAS meeting because
they have ongoing support from their school and/or community (or in a
few cases, from their new job; see Sec.~\ref{sec:jobs}). 

Some teams continue their work with their mentor scientist after their
nominal research experience year. Some alumni create new research
teams out of the alumni pool, or they just work with the ever-changing
groups of students at their schools.  There is always more demand for
alumni projects than there is mentor scientist time available. This is
not an easy problem to solve.  

While survey answers and alumni attendance indicated vaguely the
importance of links to the astronomy research community, reflection on
the importance of this (and, e.g., keeping track of numbers of alumni
attending meetings) is a more recent development.

One manifestation of these links to the wider astronomy community is
that teachers very commonly state that they want more similar
experiences, either in astronomy or in other fields. Many write
impassioned pleas for NITARP to provide more opportunities for them
personally, or to expand and reach more people in more fields.   

\begin{quote}
{\em  There is a part of me that really wants to explore ways to make the NITARP model more widely available without watering it down. I wish I'd been aware of this opportunity years ago. [...] NITARP was unique in its format and the focused way in which it works. This needs to be expanded.} -- NITARP educator, 2016 class
\end{quote}

\begin{quote}
{\em I now seek out other teachers and opportunities outside of my own school (as well as within). I am looking for collaborative research experiences so I can share the experience I have had with NITARP. While the experience will not be the same I believe it will be enough to entice other teachers to up their game as well. I have several projects in the works and each have plans of culminating with the students presenting in a professional forum.} -- NITARP educator, 2015 class
\end{quote}

\begin{quote}
{\em I do a lot of summer PD, and this was by far the best. The ongoing social/professional contact with the scientists, the new social/professional/collegial relationship with students and the expanded professional community with the other teachers are inspiring and unparallelled.} -- NITARP educator, 2016 class
\end{quote}

Many alumni are still involved with the community online.  As part of
encouraging the CoP (Sec.~\ref{sec:commofpractice}), the mailing list
is kept active. It is clear that this is valuable to alumni because
they actively seek to remain on the mailing list through job changes.

\begin{quote}
{\em Let this be your monthly installment of ``I really appreciate your emails'' [on the NITARP mailing list]. I don't know how you developed the capacity to keep up on the events in our field but I rely on you now more than ever.} -- NITARP educator, 2010 class, from an email sent in 2015
\end{quote}

Beyond the mailing list, the alumni community continues to be a link
to the research community. Staff remain available to answer any
astronomy questions, relay additional opportunities and resources,
provide `continuing education' video tutorials (see
Sec.~\ref{sec:commofpractice}) and pass along research papers with
significant results.  All of these efforts encourage connections to
the research community, as well as a culture of sharing.

\subsection{Views of Learners}

\subsubsection{The `Best Thing about the Trips'}
\label{sec:bestthingtrips}

NITARP pays for reasonable travel expenses for three different trips
during a NITARP year (see Sec.~\ref{sec:highlevelsummary} for a
description). Especially for the summer visit, the teams are kept
very, very busy for 8-9 hours per day, which (given the existent
feedback) may be the hardest/longest some students have ever worked.
Many teams add to their summer experience by sightseeing in the
evening (e.g, Griffith Observatory; Mount Wilson) or on the way out of
Los Angeles to their respective homes (e.g., the Stratospheric
Observatory for Infrared Astronomy, SOFIA, in Palmdale, CA; college
tours).  For the AAS, usually there is less ancillary sightseeing
because there is no formal NITARP-organized tour, and teachers (and
students) have to return to school, which is typically already in
session.

Despite the work load, especially since the second two trips include
students, the possibility exists that the trips could be regarded as
primarily sightseeing field trips, as opposed to the work trips they
are. In the pre-NITARP (Spitzer) years, some students were seriously
dismayed to learn that they would not be going to Disneyland or the
beach. So, even for the first four NITARP years, the feedback forms
explicitly asked, ``What was the best part of the trip?'' This is an
open-ended question; it is not multiple choice. The hope would be that
participants would not prioritize the sightseeing over the science. 

\begin{figure}
\includegraphics[width=3in]{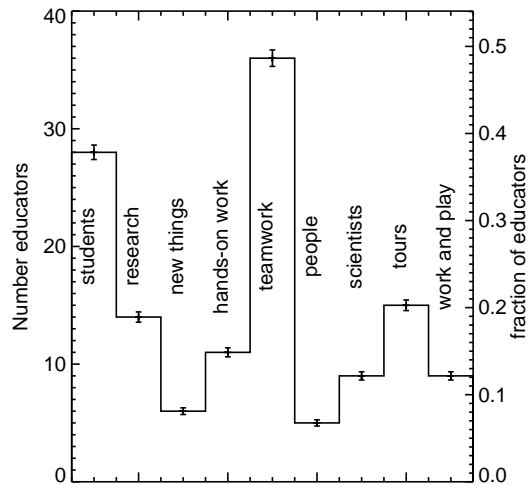}
\caption{Number (left axis) and fraction (right axis) of educators'
responses to the question, ``What was the best part of the trip?'' The
specific values are listed in Table~\ref{tab:counts}. 
students (28; 38\%);  Nearly half the teachers cite teamwork as the
best thing about the trips; see the text for additional discussion. 
\label{fig:bestthing}}
\end{figure}

The responses to the `best part of the trip' can largely be encoded
among 9 categories (see Figure~\ref{fig:bestthing}): working with
students, doing real research, learning new things, doing the hands-on
work, working as a team, meeting new people, meeting/working with
scientists, taking the tour(s), and balancing work and play.  Some
educators mention more than one item, placing them in more than one
category. The fractions for the numbers given here, in
Figure~\ref{fig:bestthing}, and Table~\ref{tab:counts} are the
fractions of educators (not fraction of responses) whose responses
place them into that category.  

Basically half the teachers cite teamwork as the best thing about the
trips (36 people; 49\%). This was not expected; the importance of the
role of teamwork has been underestimated by the NITARP organizers (see
Sec.~\ref{sec:sharing} on collaboration and sharing and Rebull
\etal~\cite{rebull2018prperpaper1} for more about the role of
collaboration in NITARP). The rate at which teamwork is listed as the
best thing about the trip often reflects that the team has, in fact,
bonded and learned a lot during the trip.

\begin{quote}
{\em Getting to be part of a team of astronomers doing science has always been a dream of mine.} -- NITARP educator, 2017 class
\end{quote}

\begin{quote}
{\em [The best thing was] Getting the opportunity to work on a project such as this with other educators from various disciplines and backgrounds come together to form friendships, working relationships, and build support for the project and each other. } -- NITARP educator, 2017 class
\end{quote}

\begin{quote}
{\em Students and teachers from all schools worked exceptionally well together.} -- NITARP educator, 2017 class 
\end{quote}

Next most popular is working with students, either their own or from
other educators (28; 38\%), which perhaps is not surprising for a
group of highly motivated teachers (also see Rebull
\etal~\cite{rebull2018prperpaper1}).

\begin{quote}
{\em My students felt more confident and better understood the big picture much more after the trip.  Nothing I could have provided would have allowed for this much growth in these two areas.} -- NITARP educator, 2014 class
\end{quote}

\begin{quote}
{\em The most interesting part of my experience was how well our student teams bonded to successfully work, and play, together.  It was amazing and an important display of cooperative learning.  They did not hesitate to help each other as well as the teachers.} -- NITARP educator, 2015 class
\end{quote}

\begin{quote}
{\em It was very surprising how quickly the students picked up the project and made great strides. They asked questions of [our mentor] and the teachers, which helped clarify information for everyone.} -- NITARP educator, 2017 class 
\end{quote}

The remaining structure in Fig.~\ref{fig:bestthing} is somewhat more
difficult to interpret. Doing real research, learning new things, and
doing the hands-on work are all specific goals the program has for the
summer visit; 26 unique educators list one or more of these categories
(some are coded with more than one of these terms), 35\% of the
sample. Meeting and working with new people and/or scientists is less
commonly cited, with 14 educators (19\%) mentioning those topics.
Finally, 23 educators (31\%) cite the tour(s) or the balance of work
and play activities as the best thing about the trip. That means that
the goals the program has for the trip (doing real research and
learning new things, via hands-on work) are cited at about the same
rate as the things the program hopes educators will not particularly
prioritize (tours and balance of work and play).  Certainly, there are
enough astronomy-related sites in Los Angeles that touring alone could
fill a complete week, but NITARP trips are supposed to be research
trips. The literature suggests students participating in short-term
``flash and dash" science activities such as visiting a museum may
positively change students' (or teachers') perceptions towards
science; there is little research that demonstrates such events bring
about long-term changes in perceptions of science or changing in
understanding of the nature of scientific inquiry
\cite{laursen2007good,csenturk2014effect}.  

Similarly, to enact changes in teachers' classrooms, PD opportunities
need to offer targeted professional development over at least 50 hours
\cite{wei2009national}.  Long-term professional development lasting
6-12 months has been linked to increased student achievement
\cite{yoon2007reviewing}. The NITARP program strives to provide such
long-term interactions, and as such, despite the obvious appeal of
sight-seeing, and despite the more mundane reality of sitting in a
room in front of a computer, touring is not a priority; learning
science by doing it should be the more influential take-away from the
trip, or at least the more influential in the long-term. However, the
results seen in Fig.~\ref{fig:bestthing} suggests that teachers list
the touring-related topics at a comparable rate as the
research-related topics when asked about the best thing about the
trip.   

\subsubsection{Comfort with the Unknown}
\label{sec:unknown}

Through the authentic research experience, educators are taught a
pattern of how to learn in a new way than they may have known before.
The experience (and the CoP) provides materials to grow with. To this
point, neither the organizers nor more generally the participants have
recognized this as one of the products of the program; this theme
emerged upon reading the surveys as part of this work. 

Participants are thrust into an environment where the learning curve
is steep. They are provided lots of active support as they learn, but
nonetheless there is a lot to learn in a very short time. For many of
these teachers, it has been a long time since they were students, and
they have work to keep up. However, by the end of the year, they are
much more comfortable with the unknown and learning new things. They
have a personalized strategy (or set of strategies) to tackle
significant new concepts or projects, or maybe just confidence that
they will figure it out, given enough time.  They are accustomed to a
`comfortable frustration' level.

It is not easy to quantify this effect based on the available data.
One way that can show evidence of this learning process is to see how
people approach their second AAS. Their first AAS is overwhelming.
But, for the second AAS, teachers know how to tackle it.  Again, this
is something that emerges from the answers they choose to provide;
$\sim$40\% of the participants wrote responses similar to the
following:

\begin{quote}
{\em EVERYTHING had a different flavor this year.  [...] I experienced everything through the lens of the research project of the past year.  The entire experience was in context. Although I was interested in seeing what the other groups had done I was far more focused on the ``what ifs'' and ``what next'' of the process.  I chose sessions based upon what I knew and how it would clarify some questions we had.  I gravitated toward posters that related to our research and asked questions of people that continually filled gaps or opened up new questions.} -- NITARP educator, 2010 class
\end{quote}

\begin{quote}
{\em This experience convinced me even further that I can push myself to learn even more each and every day. I started out feeling very overwhelmed and unprepared last year. I forced myself to work through my unease and wound up much more comfortable.} -- NITARP educator, 2011 class 
\end{quote}

\begin{quote}
{\em I now also have the tools to begin seeking out ways to expand this work on my own. I would not have had an easy start with this beforehand. [...] The skills this program provides are critical for student preparation and most teachers have not been given these. How can they then be expected to teach them?} -- NITARP educator, 2016 class
\end{quote}

\begin{quote}
{\em I feel like a popcorn kernel that has just burst open. I've grown so much at this meeting! } -- NITARP educator, 2013 class
\end{quote}

\begin{quote}
{\em By far the most interesting thing to me was the experience of presenting the poster.  Not only did it make me feel like I was really part of the conference, it made me look back to the previous AAS when presenting a poster was not just very scary, it was almost inconceivable.  So what a journey!  Reflecting on how overwhelmed and terrified I was last year, how hard I worked to learn everything, and then to be there actually doing it with some confidence, well...priceless.  } -- NITARP educator, 2012 class
\end{quote}

\begin{quote}
{\em I liked how I didn't know everything about my project to begin with. It made me become a better learner because I was asking the questions for understanding the content. In the end, I was in charge of my learning and I learned so much because of it. } -- NITARP educator, 2013 class
\end{quote}

\begin{quote}
{\em The most important thing I learned was that it's ok sometimes to not know the answer. As teachers, many times we become so consumed by having the right answer for students. Meanwhile, our students are so consumed by finding the right answer that they miss the learning. This week showed me that no matter how much work you do (in graph, periodogram, histogram, phase curve, or whatever form) you may still not come to the conclusion you thought you would... and that's ok!} -- NITARP educator, 2013 class.
\end{quote}

\begin{quote}
{\em This experience will be hard to top.  I may try to create a partnership with staff at a local university or community college to do more research projects.  I also want to get better at programming.  That is a valuable skill to share with students.  NITARP helped me to see these opportunities.} -- NITARP educator, 2016 class
\end{quote}

A second-order effect of this is that they now report being even more
frustrated than they might have been before with traditional PD
opportunities that are available.

\begin{quote}
{\em I can say that my expectations for professional development in the future are such that I will not be satisfied with most opportunities that are offered locally. Thus, I feel that I will take the opportunity to offer PD to other teachers, particularly as it pertains to the nature of science or how to conduct research with archived data.} -- NITARP educator, 2016 class
\end{quote}

\begin{quote}
{\em  No district-led professional development can compare [to NITARP].   I am very excited to lead some professional development opportunities in my district, but that is only a very small slice of the pie about NITARP.  [...]I am looking actively to find other integrated professional development opportunities for teachers which make us step out of the classroom and work with professionals in the community.} -- NITARP educator, 2014 class 
\end{quote}

\begin{quote}
{\em The BEST subject-area professional development experience I've had in 25 years BY FAR, and one of the most intellectually stimulating experiences I've had in years. I lie awake at night thinking about data.} -- NITARP educator, 2014 class 
\end{quote}

\begin{quote}
{\em I have said this many times and will continue.  I have been a teacher for 38 years, and have been in probably 18-20 special programs over that time to improve myself as a science teacher.  The NITARP program ranks as one of the three best programs I have been in over that period of time.  } -- NITARP educator, 2013 class 
\end{quote}

\subsubsection{Student Empathy}
\label{sec:studentempathy}

NITARP requires a lot of its educators and students. Even those who
begin participation knowing that they will need to work hard are
frequently overwhelmed.  For many educators, it has been a long time
since they have been in the position of a learner, especially one
struggling to understand; they admit that they have forgotten what it
is like to be a student.  Some educators (about 7\% of our sample)
explicitly note, without being prompted to do so, that as a result of
their experience, they now have much more empathy for their
overwhelmed students. While the occasional instance of this in
feedback forms had been noted, combing feedback forms as part of this
work made this theme crystallize better than it had before.

It is notable that these statements of student empathy come most
commonly, but not solely, from educators who are often the most
overwhelmed educators, those who, at first appraisal, may seem to not
have gotten much out of the experience. Increased student empathy is a
clear benefit to those educators (and their students), even if they
did not understand all the rest of the experience to the depth one
might wish.

\begin{quote}
{\em I got to experience what it's like to be a student struggling with exciting new material. This has increased my functional empathy with students.} -- NITARP educator, 2016 class 
\end{quote}

\begin{quote}
{\em This experience will help me understand how students feel when they are presented with new material and don't understand.  I think this will give me more patience and understanding in this area.} -- NITARP educator, 2012 class 
\end{quote}

\subsection{Pedagogical Changes and Professional Growth }

\subsubsection{Professional Growth}
\label{sec:jobs}

As people develop professionally, they look for new experiences. Some
people discover NITARP because they are specifically looking for new
opportunities.  At least 80\% of participants report that they are
actively seeking opportunities to learn, grow, and change as part of
NITARP and that the program meets that need.

Some participants change career paths rather dramatically during or
after their participation,  often into positions that allow them
greater influence over more educators and/or policy. While, of course,
the program can't claim credit for all or even most of that, it may
have facilitated it. Their experience may have opened the educators'
eyes to different career paths than they might have previously known.
This theme emerged from the first four years of NITARP (and the prior
years with Spitzer), and was explicitly probed with survey questions
for the second four years.

From alumni data, it is suspected that the program had a significant
role in the career changes of at least 9 alumni ($\sim$12\% of all
participants); however, records are incomplete. At least two of those,
plus 2 more alumni, started graduate school in education at least in
part as a result of their NITARP experiences.  At least one more
specifically mentioned her mentor educator experience as critical to
her realization that she would prefer to teach teachers than be in her
(then) current high school classroom. Several people moved up in their
district's hierarchy such that they can teach teachers or set policy
at a district (or higher) level; this is a positive outcome in that
the influence of NITARP can then be extended to a larger community
than would be possible had those teachers stayed in the classroom.
Some now work for observatories or large astronomy projects as part of
education efforts. 

\begin{quote}
{\em [...] I'm now a Science Instructional Coach who works mainly with teachers and I'm able to impart accurate information about what scientists really do to middle school science teachers who don't really know what that is.  I'm able to help them design lessons and science fair projects that allow students to experience authentic research activities.} -- educator, 2005 class, writing in 2013
\end{quote}

\begin{quote}
{\em [...]my NITARP experience made my science department realize that we need to bring the use of real data into our curriculum.  Since we are reorganizing because of the new science standards, it is an opportunity for us to do this.  We realize the support for our students to handle large data sets will need to be scaffolded.  Having worked with NITARP doing archival research I am now working with my dept. chair to bring a research component into all our science classes.  The experience that I had with NITARP was so inspiring that I am more than willing to donate my time for this.} -- NITARP educator, 2012 class, writing in 2013
\end{quote}

It is important to note that there is significant selection bias
likely to be affecting these results. Educators who apply to
participate are already looking for new opportunities to
learn/grow/change; in some sense, then, it is not surprising that a
significant fraction change jobs. However, the magnitude of the
changes, and the direction of the changes, in at least a few cases,
are larger than would be expected for a highly capable classroom
educator. The changes are attributed by the educators themselves to
their NITARP experience, in part or in whole.

Because significant career change has emerged as a trend, participants
are now probed more explicitly about professional goals, but this is
still in the nascent stage. Additionally, there will need to be more
complete and explicit surveying of alumni on longer timescales than
the primary year to complete this kind of analysis; more time is
needed to explore the ramifications of the program on long-term
professional growth. Moreover, the effects on mentor educators (who
spend several years on active teams) maybe be significantly different
than participant educators (who spend only one year on a team).

More broadly, 14\% of alumni used the words `life changing' when
describing the impact of their research experience on their lives. 
NITARP provides PD at the level that high-achieving teachers need.

\begin{quote}

{\em  My NITARP experience has made me rethink my entire approach to science education. Many of my students expect me to do the work and pretty much hand it to them all wrapped up and neat. Science education must involve a great deal of discovery by the student and not a string of topics with definitions. } -- NITARP educator, 2012 class 
\end{quote}

\begin{quote}
{\em My life has changed in some way because of my participation in this program.  My wife, my children, and my co-workers have all remarked at how I am different now.  I don't know whether it was the program, the people that I worked with, or some combination of the two, but whatever it was something about it changed me.  I know that ``life-changing experience'' was not one of the outcomes that you hoped for when you planned the program, but it is what happened with me. Thank you very much for allowing me to participate -- this has been one of the best years of my life.} -- NITARP educator, 2013 class
\end{quote}

\subsubsection{Better Science in Their Classrooms and Their PD}
\label{sec:betterscience}

Information can be teased out about how more authentic science might
be making its way into educators' classrooms and PD experiences that
they lead. In one case, prior to their experience, one educator was
proud of the `real research' he was helping students conduct at his
school. After his experience, he reported that he realized that since
he knew what answers the students would find before they started, that
it clearly was not real research. One assumes that he then
subsequently made changes to address this in his classroom, but that
is an assumption.

The survey results, since they were collected during and immediately
after the intensive NITARP experience, often reflect a realization on
the educators' part that they need to change what they are doing in
the classroom. Surveys over a longer time baseline covering what
changes were actually made in the classroom (as opposed to changes
they want to make) are beyond the scope of the present work. From
examining survey results like this, though, at least 60\% of the
participating educators report that they want to bring more authentic
(`better') science into their classrooms and into the PD they lead
following participation in the program. 

This theme emerged in the first 4 NITARP years and then was explicitly
probed in the second 4 NITARP years.

\begin{quote}
{\em [NITARP] has made me realize that while I use a lot of inquiry, I don't always involve my students in the process of developing a testable question.  [...] The next time, I hope to involve my students more in the entire process.  I plan on emphasizing that science is a collaborative effort.} -- NITARP educator, 2012 class  
\end{quote}

\begin{quote}
{\em After this experience, I'm more aware that beyond just teaching my students good science, my focus should be to prepare them for a career in science. I feel like I have a better understanding of the skills they'll need to be successful. I've already made big changes to my curriculum because of this program, and will continue to do so in future years.} -- NITARP educator, 2012 class 
\end{quote}

\begin{quote}
{\em All of this is directly applicable to my classroom.  What is it that makes some learners want, need, desire a cookbook style while others would rather discover for themselves?  In a traditional classroom situation the learner who enjoys the freedom of less direction does not always fit the teacher's mold.  This student might go off on interesting tangents as they investigate the topic at hand which isn’t always easy to manage in a class of 30 students.  Team work does not necessarily mean everyone is working hand in hand, but instead that the group is working towards the same end product, with each individual finding their own way at times.  Individual work and subsequent sharing allows learners to leapfrog  over each other towards the end product.  
So the question is, how do we balance direct instruction and open ended inquiry?  How do we make the average learner more comfortable with the open ended approach?   How do we pull away from ``cookbook'' learning and labs and free our learners to investigate the many possible paths to the end product?  How much direct instruction is necessary and at what point to we leave the learner to their own devices to ``figure things out?''  I need to find more ways to make this happen in my own classroom. } -- NITARP educator, 2012 class 
\end{quote}

\begin{quote}
{\em My science department is already discussing how to bring in and scaffold both programming and use of existing data bases in all of our science classes as a result of my participation in NITARP and the demand of our students, based on feedback from our alumni.} -- NITARP educator, 2012 class
\end{quote}

\begin{quote}
{\em I have a new view on doing research. I now understand the need for solid science programs at all levels as a foundation for future scientists, not just astronomers. This experience has made me reflect on my pedagogy and motivates me to continue to strive to improve my techniques.  } -- NITARP educator, 2013 class
\end{quote}

\begin{quote}
{\em NITARP has made me realize that most science teachers don't really immerse students in real science. Too often we (myself included) do `labs' that have an answer and fit in a class period, and though NITARP has made me want to deviate from this, I'm still not exactly sure how yet. All I know is my students don't ask as many questions as they could be and should be and I need to work on it.  } -- NITARP educator, 2013 class 
\end{quote}

\begin{quote}
{\em I am more compelled than ever by the collaborative model of doing research with high school students. I also feel strongly that it is wonderful to expose students to the public archives.  I will continue to work on exposing my introductory and advanced students to these two aspects of doing research.} -- NITARP educator, 2014 class
\end{quote}

\begin{quote}
{\em In our research course at school, I am pushing to change some of the parts of it that need updating and that don’t really reflect the reality of scientific work. For instance, assembling a complete research proposal, which now seems so obvious, was not a requirement at school. I am also planning on providing students with opportunities to conduct research where they do not need to collect their own data. Why not? There's so much already out there. } -- NITARP educator, 2016 class 
\end{quote}

However, the ability to assess the degree of more/better science in the classroom is strongly affected by not only the questions asked of the educators, but how they were asked (and most likely what was discussed with the educators on any given team over the year). The literature based on other teacher research PD suggests that such changes are likely \cite{westerlund2002,blanchard2009,houseal2014impact,laursen2007good,westerlund2002}. However, this topic specifically in the context of NITARP requires more work to fully understand, in particular both better questions and better post-NITARP follow-up to probe the longer-term impact on educators, and specifically how NITARP experiences influence classroom behavior on both the short and long term.

\subsection{Limitations}
\label{sec:limitations}

The data used here is entirely self-reported data from a relatively
small number of teacher participants, albeit highly capable teachers
who can recognize changes in their approaches and/or philosophies.
Moreover, the program is selective and the teachers highly motivated.
Recurrent themes are triangulated between multiple surveys from the
same person at different waypoints and multiple surveys at the same
waypoint from different people. Increasing the sample size by simply
waiting for more years to pass is possible, though NITARP itself, and
the questions that are asked at the waypoints, change with time as the
program continually readjusts to meet new needs. Longer time baseline
studies of the same people over time are also desirable, to sample,
for example, changes made in the classroom in response to NITARP.

Another limitation is that the data from the first four years used
here often included incomplete or missing answers to questions
(because of the survey design; see Rebull
\etal~\cite{rebull2018prperpaper1}). As such, any information on some
of the themes investigated here is not available (or not as clearly
revealed) for those earlier participants.  

Because some of the themes identified here emerged in the process of
conducting this research, these themes were not explicitly probed with
questions in the surveys.  Interviews with participants specifically
on the themes investigated here could also illuminate the themes
investigated here. Investigating similar themes in student data is
beyond the scope of this work.

This work does not track any changes in classroom instruction; it
works primarily with teachers' perceptions and not how or when these
perceptions impacted the classroom, beyond teachers explicitly stating
that they were rethinking their approach. Classroom impact is left to
future work.


\section{Summary and Future Work}
\label{sec:summary}

Despite the fact that many educators did not participate in authentic
science research before being in the classroom, changes to science
education in the US continue \cite{standards2013next,naframework}. As
educators are asked to include more authentic science experiences in
the classroom, there is increasing demand for PD that fills that gap. 
Since educators tend to teach in the same way in which they were
taught
\cite{brown2006college,crawford2014,duschl2007learning,quinn2012national},
if they have never conducted authentic science research, it can be
very hard for them to teach students how to be engaged in scientific
practice.  

NITARP, the NASA/IPAC Teacher Archive Research Program, partners small
groups of mostly high school classroom teachers with a research
astronomer for a year-long authentic astronomy research project.
Operating since 2005, by 2017, the program has worked with a total of
103 educators from 34 states. The empirical data used in the
qualitative  analysis here focuses primarily on the last eight years
(2010-2017) of surveys collected at up to 4 waypoints from 74 educator
participants. 

The original research question was: how do educator participants
describe the major changes and outcomes in themselves fostered by
NITARP?  This is a summary of self-report evaluation data, though only
about half of the themes discussed here were specifically targeted
with survey questions. 

Evidence was found that the program helps foster an array of skills in
the participants. Following the program, educators have a more
accurate view of the nature of science; 74\% of teachers report some
or major change in their understanding as a result of the program.
Educators emerge with a better understanding of who astronomers are
and qualities they possess; $\sim$60\% list patience and/or
persistence as important, $\sim$40\% mention creativity, and
$\sim$30\% list collaboration and/or teamwork. The program fosters
collaboration and sharing within their teams, and enables better
subsequent sharing with other educators and scientists. The program
provides an ongoing strong link to the astronomy research community. 
It helps educators to have more comfort with the unknown, and greatly
increases their personal confidence in their ability to learn entirely
new skills.  At least 7\% of participating educators report increased
student empathy. The program fosters professional growth; at least
80\% of educators say they participate because they are looking to
learn, grow, and/or change. At least 12\% have changed their career
path significantly, based in part because of their NITARP experience.
Moreover, 14\% of the alumni say the experience was life changing. 
And finally, the program enables more authentic science methods to be
modeled, demonstrated and expected in the classroom; at least 60\% of
alumni self-report that the program changed what they want to do in
the classroom. 

The majority of research on science teachers' research experiences has
focused on content knowledge gains, perceptions of science and
scientists, changes in pedagogy, and teachers' interactions with
scientists
\cite{blanchard2009,dresner2006teacher,houseal2014impact,laursen2007good,raphael1999research,westerlund2002}.
Teachers' frustrations with working through science research projects
were noted in Burrows \etal~\cite{burrows2016authentic}. There has
been only limited systematic work on teacher research experiences
\cite{buxner2014exploring,sadler2010}. This work adds to the existing
body of literature to illuminate what teachers themselves say they
have gained from the experience. 

The results, analysis, and conclusions we make in this study are
generalizable to other teacher research experience programs and, more
importantly, open up more in-depth research questions for future
exploration. Future research specifically can be envisioned into the
impact of NITARP and similar PD programs. For example, it would be
interesting to probe the effects on the student participants and on
the students of participating educators in years subsequent to the
intensive research experience year. The long-term impact of NITARP
should be investigated, including specifically the impact on mentor
educators, who spend multiple years in the program.

At the current funding level, only a few teachers per year can be
involved in NITARP.  Ideally, more teachers should be involved, over
more science subjects. But until that goal is realized, NITARP alumni
share their experience of what actually doing scientific research is
really like with other educators who have no such experience. Sharing
of their experience is required, specifically because so few educators
can participate per year. The teachers who are reached only through
the NITARP alumni could obviously learn more if they participated in
NITARP, but, given the lack of resources, at least learning,
peer-to-peer, about the nature of science, is a step in the right
direction. For these second-order teachers who then want to begin by
incorporating more data into their classrooms, they learn that
archival data are available and that there are  programs that
incorporate real data into the classroom, even if not to the same
rigor expected of NITARP participants (see, e.g., Rebull
\cite{rebull2018rtsrekeynote}).

The NGSS calls for teachers to implement more authentic practices in
their science classrooms. Because many teachers lack the requisite
training, PD programs such as NITARP are going to play an important
role at helping teachers gain the knowledge and skill to implement the
NGSS with fidelity.  NITARP shows that these kinds of experiences are
instrumental at changing teachers' attitudes and behaviors in the
classroom.  These results show that teachers are capable of doing
science research and implementing authentic science in their
classrooms. This work suggests that more PD programs should involve
authentic research activities as part of their program. Such PD will
first meet the higher-level needs of high achieving teachers; later,
programs such as this could be expanded to provide the infrastructure
enabling all teachers to be ready to engage in authentic research, and
then support them in that endeavor.  If PD providers have high
expectations for their teachers, then those expectations can be met,
just as high expectations for students can be met
\cite{marzano2007art}.

\begin{acknowledgments}
Thank you to all 103 NITARP \& Spitzer educators for your tireless devotion to this program! 

Support for this program was provided in part by NASA/ADAP funds. Thank you Doug Hudgins!

Thanks to Martha Kirouac and Tim Spuck for useful suggestions on early drafts.
\end{acknowledgments}


\bibliography{References.bib}

\end{document}